\documentclass[fleqn,usenatbib]{mnras}

\usepackage{newtxtext,newtxmath}

\usepackage[T1]{fontenc}

\DeclareRobustCommand{\VAN}[3]{#2}
\let\VANthebibliography\thebibliography
\def\thebibliography{\DeclareRobustCommand{\VAN}[3]{##3}\VANthebibliography}

\usepackage{graphicx}	% Including figure files
\usepackage{amsmath}	% Advanced maths commands

\usepackage{siunitx}
\usepackage{booktabs}
\usepackage{gensymb}
\usepackage[table,xcdraw]{xcolor}
\title[The solitary star cluster of And~XXV]{The solitary star cluster of the Andromeda XXV dwarf spheroidal}
\author[C. Crociati et al.]
{C. Crociati,$^{1}$\thanks{E-mail: ccrociat@ed.ac.uk}
A. M. N. Ferguson,$^{1}$ G. McGill,$^{1}$ R. Pascale,$^{2}$ A. Genina,$^{1}$ P. B. Kuzma,$^{1}$ \newauthor  D. Mackey,$^{3}$, A. W. McConnachie$^{4}$ and R. \v{Z}emaitis$^{1}$
\\
$^{1}$Institute for Astronomy, University of Edinburgh, Royal Observatory, Blackford Hill, Edinburgh EH9 3HJ, UK  \\
$^{2}$INAF - Osservatorio di Astrofisica e Scienza dello Spazio di Bologna, via Gobetti 93/3, 40129 Bologna, Italy \\
$^{3}$Independent Researcher, Charnwood, Canberra, ACT 2615, Australia\\
$^{4}$National Research Council Herzberg Astronomy and Astrophysics, 5071 West Saanich Road, Victoria BC V9E 2E7, Canada}

\date{Accepted 2026 August 28. Received 2026 August 24; in original form 2026 June 15}

\pubyear{\the\year{}}

\newcommand{\Rh}{R_h}
\newcommand{\FeH}{\rm [Fe/H]}
\newcommand{\bbtheta}{\boldsymbol{\Theta}}
\newcommand{\dd} {{\text d}}
\newcommand{\Area}{\mathcal{A}}

\definecolor{lightgray}{gray}{0.9} 

\begin{document}
\label{firstpage}
\pagerange{\pageref{firstpage}--\pageref{lastpage}}
\maketitle

% Abstract of the paper
\begin{abstract}
%The abstract should briefly describe the aims, methods, and main results of the paper.
%It should be a single paragraph not more than 250 words (200 words for Letters).
%No references should appear in the abstract.
%Modelling the formation and survival of globular clusters (GCs) in very low-mass galaxies represents both an empirical and theoretical challenge deeply linked to our understanding of galaxies formation and cosmological models. It is therefore imperative to confirm and characterise GCs candidates in tiny galaxies, given that only a handful of cases have been observed so far. 
We present {\it Hubble Space Telescope} Advanced Camera for Surveys observations of Gep~I, a globular cluster (GC) candidate in the low-mass ($M_{\star}\sim 6.5 \times 10^5\,M_{\sun}$) M31 dwarf spheroidal (dSph) satellite Andromeda~XXV (And~XXV). We confirm the nature of this object and provide the first detailed characterisation of its resolved stellar populations using a colour-magnitude diagram (CMD) that reaches 2 magnitudes below the horizontal branch. We compare Gep~I's metallicity and distance with those of the surrounding And~XXV stellar population, and find them to be strikingly similar, consistent with a physical association between the GC and the dSph.  Gep~I is very extended ($\Rh=24^{+5}_{-4}$ pc) and faint ($M_V = -4.5 \pm 0.2$ mag), similar to the star clusters residing in other low-mass dwarf galaxies. 
It is characterised by a very low metallicity ($\FeH = -2.4^{+0.3}_{-0.4}\,$dex) and a red horizontal branch morphology, a combination also seen in suspected accreted GCs in the M31 halo and in the Local Group dwarf irregular galaxy NGC\,6822. While Gep~I is mostly likely a genuine star cluster, the current data do not exclude the tantalising possibility that it consists of And~XXV stars temporarily captured by a dark subhalo orbiting within the dSph's potential well.
%However, this scenario would be at odds with the old age of Andromeda~XXV, whose estimated quenching epoch is $\geq$10 Gyr. 
\end{abstract}

% Select between one and six entries from the list of approved keywords.
% Don't make up new ones.

%%%%%%%%%%%%%%%%%%%%%%%%%%%%%%%%%%%%%%%%%%%%%%%%%%

%%%%%%%%%%%%%%%%% BODY OF PAPER %%%%%%%%%%%%%%%%%%

\section{Introduction}

Globular clusters (GCs) residing in the smallest galaxies are key to understanding how star formation and hierarchical assembly proceed in low-mass systems.  While clear correlations between GC system mass and host galaxy macro-properties -- such as the dark matter (DM) halo and total stellar mass -- are well-established in the high-mass regime, the behaviour of these scaling relations in the dwarf-galaxy regime remains remarkably uncertain \citep[e.g.][]{georgiev+2010, forbes+2018, eadie+2022, berek+2024, dornan+2026}. 
Furthermore, both the frequency of GCs in low-mass galaxies, as well as their typical properties, are still poorly constrained. These deficiencies in our understanding mean that theoretical models of the formation and evolution of stellar clusters in dwarf galaxies lack crucial observational constraints \citep[e.g.][]{el-badry+2019,chen_gnedin2023, doppel+2023, gutcke2024low-mass-1dc, taylor2025emergence-63d}. 
%As a consequence, theoretical models predicting the formation and fate of stellar clusters in dwarf galaxies (e.g., \citealt{el-badry+2019, bastian+2020, valenzuela+2021,chen_gnedin2023}) struggle to converge on which underlying physical and cosmological prescriptions to adopt to reproduce the observed scattered trend at the smallest scales. 

In addition, observations of GCs in dwarf galaxies can provide important insight into the still-debated nature of DM. In fact, DM-only structure formation simulations in the standard cold DM cosmological framework ($\Lambda$CDM) predict DM haloes to have density profiles characterised by steep inner cusps \citep{navarro+1996}. However, observations of local dwarf galaxies favour, in many cases, a constant-density inner core (see, e.g., \citealt{Moore1994, adams+2014,battaglia&nipoti2022}). There is growing evidence (e.g., \citealt{read+2019,bouche+2022}) that stellar feedback and gas inflow can transform DM cusps into cores, but these processes can be inefficient in very low-mass, gas-depleted dwarfs (e.g., \citealt{penarrubia+2012, dicintio+2014, onorbe+2015, orkney+2021}).
The smallest galaxies are therefore optimal targets for investigating the pristine nature of DM haloes, and the presence of GCs in these systems may provide important clues. For example, several studies argue that the survival, over many Gyrs, of stellar clusters in the central regions of very low-mass dwarfs requires a central DM core \citep[e.g.][]{goerdt+2006, cole+2012,amorisco2017,contenta+2018,webb&vesperini2018, 2019MNRAS4852546B}, with further constraints coming from the GC properties, such as their ellipticity and size \citep[e.g.][]{orkney+2022}.  The more examples we can find of GCs in low-mass galaxies, the tighter these constraints on DM properties can become.  

\begin{figure*}
    \centering
    \includegraphics[width=0.9\textwidth]{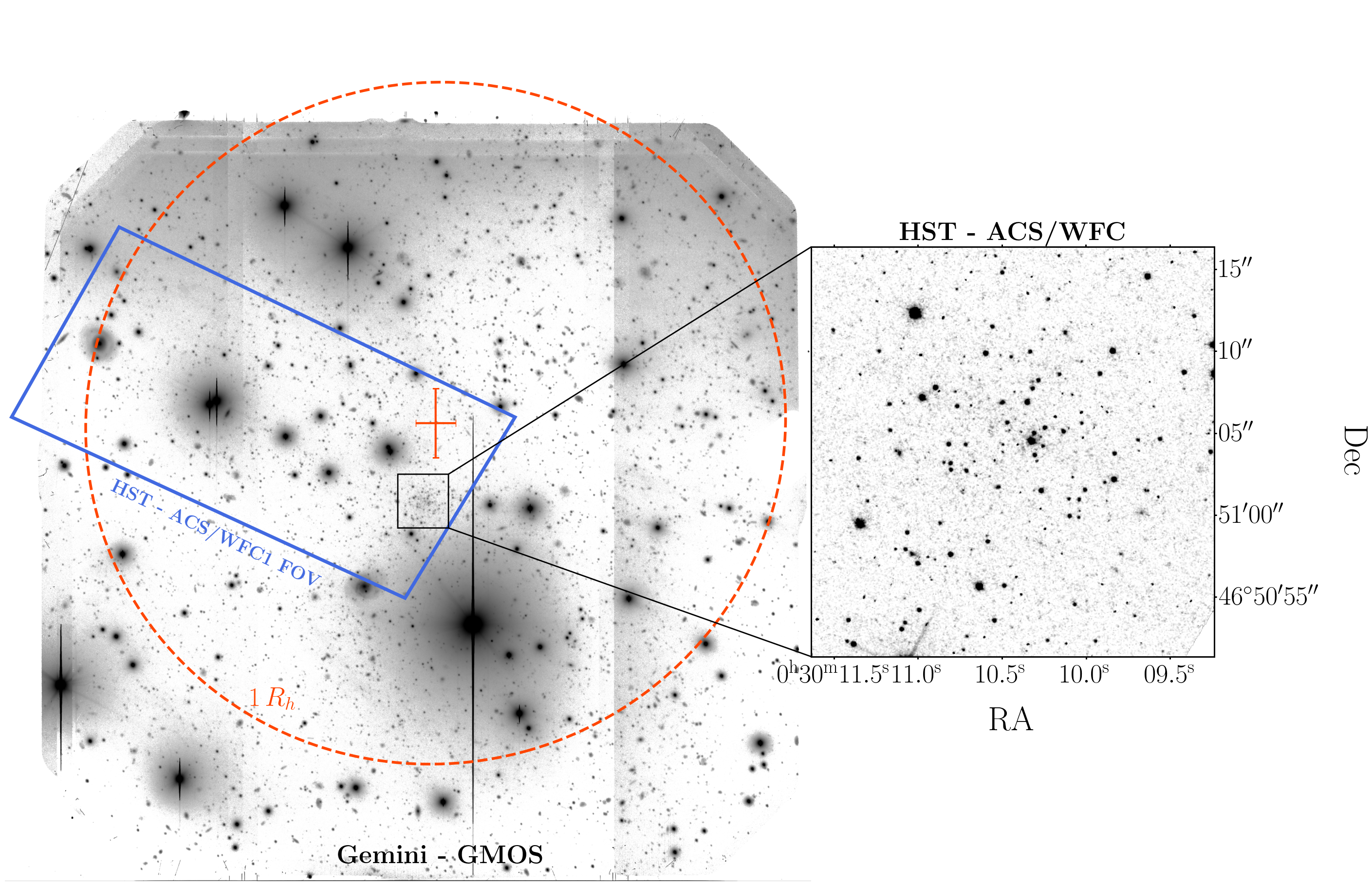}
    \caption{Stacked GMOS image in the $g$-band of And~XXV. The centre, with its associated uncertainties, and the half-light radius of the galaxy are shown in red, taken from \citealt{martin+2016}. The HST-ACS/WFC pointing capturing the stellar cluster is shown by the blue polygon. The inset panel shows a $25 \times 25\,\rm arcsec^2$ thumbnail of the HST-ACS/WFC \texttt{drc} image in the F814W filter centred on the cluster Gep~I.}
\label{fig: FOV_image}
\end{figure*}

Excluding the Fornax and Sagittarius dwarf spheroidal (dSph) galaxies, which are the only two Local Group dSphs hosting a system of GCs, only four very low-mass\footnote{For comparison, the stellar masses of Fornax and Sagittarius, as listed in \cite{forbes+2018}, are $M_{\star} = 2.0 \times 10^7\,M_{\sun}$ and $1.3 \times 10^8\,M_{\sun}$, respectively.} ($M_{\star} \leq 10^6\,M_{\sun}$) dwarf galaxies are known to host a  confirmed or candidate GC. Arranged in order of increasing stellar mass, these systems are Ursa Major~II, Eridanus~II, Andromeda~XXV, and Andromeda~I (see \citealt{forbes+2018, eadie+2022} and references therein).  Except for Eridanus~II, for which both kinematic \citep{zoutendijk+2020} and photometric (\citealt{simon+2021,weisz+2023}) investigations have been performed, the lack of deep resolved photometry and/or kinematical data hampers a firm conclusion about the nature of the remaining clusters. Here, we focus on the candidate GC of Andromeda~XXV (hereafter And~XXV), a low-mass satellite ($M_{\star}\sim 6.5 \times 10^5\,M_{\sun}$; \citealt{forbes+2018}) of M31 located at a 3D galactocentric radius of $\sim$85 kpc \citep{savino+2022}. This object, named Gep~I, was discovered by \cite{cusano+2016} using deep Large Binocular Telescope imagery acquired under in modest seeing conditions ($\sim 0.8-1$ arcsec). Although these data were sufficient to allow an estimation of the structural properties of Gep~I from integrated light, only a few red giant branch (RGB) stars were resolved and it was not possible to construct a detailed colour-magnitude diagram (CMD). No other photometric or spectroscopic studies of this cluster candidate have been published since the discovery paper. As a consequence, the nature of Gep~I has remained ambiguous until now, with no robust measurements of its stellar populations or its distance. Without such measurements, even the association of Gep~I with And~XXV is insecure, as the object could simply be a member of the M31 GC halo population seen in projection. 

Leveraging observations from the {\it Hubble Space Telescope (HST)}, we aim to investigate the nature of Gep~I and put its association with And~XXV on firmer footing. This will be carried out by performing a comparative analysis of the properties of Gep~I and its surrounding field population using deep sub-horizontal branch (HB) photometry. In Section~\ref{sec: observation_reduction} we describe the observations and data reduction, and in Section~\ref{sec: CMD_analysis} we analyse and compare the CMDs. Section~\ref{sec: discussion} is dedicated to the discussion of the results, while Section~\ref{sec: summary_conclusions} summarises our findings and conclusions.

\section{Observations and data reduction} 
\label{sec: observation_reduction}

And~XXV was observed as part of the {\it HST} Treasury Survey of M31 satellite system (GO-15902; PI: D. Weisz). Through deep exposures, this program produced deep photometric catalogues that reach to the old main-sequence turn-off for many M31 satellites \citep{savino+2022,savino+2025}.
In the case of And~XXV, only a subset of the dithered observations captured Gep~I, and therefore the published catalogue of And~XXV does not include stars lying at its location.
In this work, we downloaded the four Advanced Camera for Surveys/Wide Field Channel (ACS/WFC) images which contained Gep~I from the Mikulski Archive for Space Telescopes (MAST) archive.  Specifically, these consisted of two dithered exposures in each of the $F606W$ and $F814W$ filters, with total exposure times of $2020\,$s and $2460\,$s, respectively. 

Figure~\ref{fig: FOV_image} shows an archival stacked $g$-band image of And~XXV, obtained with the GMOS-N instrument on the 8\,m Gemini North Telescope (program ID: GN-2018B-FT-201; PI: McConnachie). The {\it HST} field analysed in this work is shown with the blue polygon, and the position of Gep~I is marked by the grey square. For reference, the galaxy centre and half-light radius, as derived by \cite{martin+2016} using data from the Pan-Andromeda Archaeological Survey (PAndAS; \citealt{mcconnachie+2018}), are also indicated.  The right panel of Fig.~\ref{fig: FOV_image} presents a zoomed-in view of the cluster in the $F814W$ \texttt{drc} image, corresponding to an area of $25 \times 25\,\rm arcsec^2$.

The photometric reduction was carried out via \texttt{DAOPHOT II} \citep{stetson1987}. We performed PSF fitting on the single \texttt{flc} images, after the correction for the Pixel Area Map had been applied. We focused only on the chip containing the cluster, i.e. chip~1. A specific PSF model for each image was determined by selecting more than 50 unsaturated, bright and isolated stars across the entire chip, and then applied to flux peaks at $3\sigma$ from the local background by means of \texttt{DAOPHOT/ALLSTAR}. The great majority of cosmic rays were rejected at this stage of the reduction by geometrically matching the first single-frame catalogues with  \texttt{DAOPHOT}'s ancillary routines \texttt{DAOMATCH} and \texttt{DAOMASTER} \citep{stetson1993}, and retaining only sources detected in both filters. This process also generated the input master star list for \texttt{ALLFRAME} \citep{stetson1994}, which provided the photometry used throughout this paper.
%, which returned single-frames final star lists. The two outputs per filter were then combined together using, once again, \texttt{DAOMATCH} and \texttt{DAOMASTER}, determining for each star its final instrumental magnitude. 
%and error, defined as weighted mean and standard deviation.  
Instrumental magnitudes were calibrated onto the VEGAMAG system adopting the zero-points and encircled energy corrections provided for the ACS camera\footnote{\url{https://www.stsci.edu/hst/instrumentation/acs}.}. Instrumental coordinates were reported on the absolute World Coordinate System by cross-matching our final catalogue with the one presented in \cite{savino+2025}\footnote{The catalogue can be downloaded from the MAST HLSP repository: \url{https://archive.stsci.edu/hlsp/m31-satellites}.} through the software \texttt{CataXcorr}\footnote{Part of the astronomical software package \texttt{CATAPACK} developed by P. Montegriffo at INAF–OAS. The software is publicly available online: 
 \url{http://davide2.bo.astro.it/paolo/Main/CataPack.html}.}. %
%This operation allowed us also to check for possible magnitude calibration offsets with respect to \cite{savino+2025}, finding a median offset smaller than 0.015 mag (rms=0.00) for stars brighter than the main-sequence turnoff in both filters. 

We culled the photometric catalogue from badly fitted and spurious sources by applying an iterative $3\sigma$ clipping cleaning procedure on the distribution of photometric errors, chi and sharpness parameters as a function of magnitude. A visual inspection of the stars surviving this cleaning process on the stacked \texttt{drc} images confirmed that sources associated with background galaxies and saturation spikes were successfully excluded.  

Our analysis focuses largely on a differential analysis of the Gep~I stellar populations with those of the surrounding And~XXV field. As crowding is not severe in either case, we assume these have similar completeness levels, and the comparative nature of the measurements means that this does not need to be determined in an absolute sense. 
%Nevertheless, considering comparable $HST$ data for M31 halo GCs analysed by \citet{mcgill+2025}, we expect to have a 50 per cent completeness level of $\sim27.5$~mag in $F606W$, roughly two mags below the HB.

\begin{figure}
    \centering
    \includegraphics[width=0.35\textwidth]{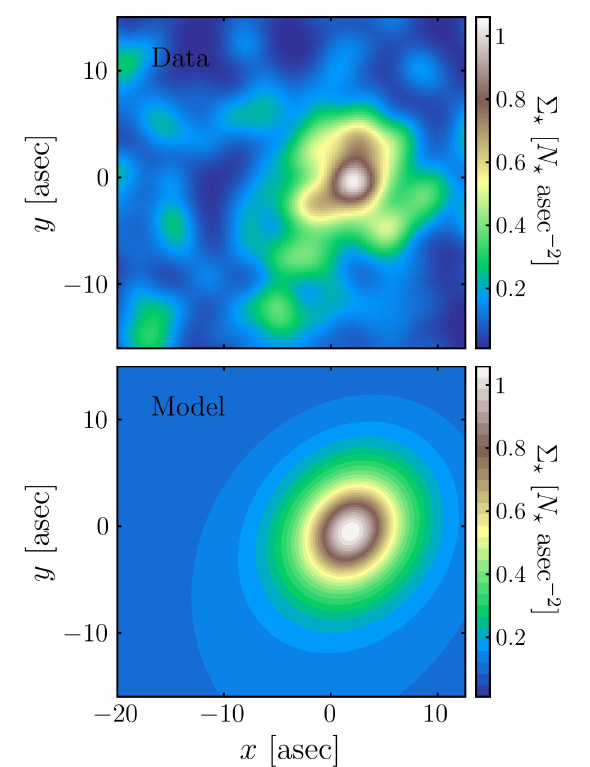}
    \caption{2D density histograms ($N_{\rm stars}/ \rm asec^2$) representing the observed distribution of stars (upper panel) and the model (lower panel). The spatial bin is equal to $0.5\arcsec$ both in x and y. The resulting histogram of observed stars has been smoothed using a Gaussian kernel with a dispersion of $1.5\arcsec$. Stars' positions are centred on the cluster centroid determined by \citet{cusano+2016}. }
\label{fig: density_map}
\end{figure}

\section{Results}
\label{sec: results}

\subsection{Structural properties}
\label{sec: structure_fit}

%As first step, we determined the characteristic radius of the cluster to distinguish its stars from the And~XXV galaxy.
The 2D map of stars in a small region centred on Gep~I is shown in the upper panel of Figure~\ref{fig: density_map}. The map, centred on the cluster's centroid as determined by \cite{cusano+2016}, was constructed using a pixel scale of $0.5\arcsec$ and subsequently smoothed with a Gaussian kernel with a dispersion of $1.5\arcsec$. The stars were filtered following the photometric quality cuts described at the end of Section~\ref{sec: observation_reduction}. In addition, stars with colours $m_{\rm F606W} - m_{\rm F814W} > 1.5$ were excluded from the analysis.

%We fitted the spatial distribution of these stars with the flattened Plummer profile \citep{plummer1911}. The fitting procedure closely follows the method outlined in \citet[see also \citealt{smith+2023,pascale+2026}]{martin+2016}, which fits individual stars in a Bayesian statistical framework and determines the probability of membership of a single star to the cluster. We considered the area circumscribed by the blue polygon in Fig.~\ref{fig: FOV_image} as the sky region, a linear gradient to model the background, and uniform priors. We employed \texttt{emcee} \citep{foreman-mackey+2013} to sample the posterior distribution of the model parameters. The lower panel of Figure~\ref{fig: density_map} illustrates the best-fit model, while Table~\ref{tab: total_tab} lists the 50th percentiles of the marginalised probability distribution functions of the model parameters, together with the 16th and 84th percentiles. In this way, we determined the cluster's central coordinates, half-light radius ($\Rh$), ellipticity ($e$), and position angle ($\phi$). In addition, we report the number of stars of the {\it HST} catalogue associated with the cluster ($N_{\star}$). %We note that the posterior distributions are well defined, with no evident correlations between the parameters of the Plummer profile. 

We fitted the spatial distribution of these stars with the flattened Plummer profile \citep{plummer1911}. The fitting procedure closely follows the method outlined in \citet[see also \citealt{smith+2023,pascale+2026,Bellazzini2026}]{martin+2016}. We considered the area circumscribed by the blue polygon in Fig.~\ref{fig: FOV_image} as the sky region, a linear gradient to model the background, and uniform priors. In this way, we determined the cluster's central coordinates, half-light radius ($\Rh$), ellipticity ($e$), and position angle ($\phi$). In addition, we report the number of stars in the {\it HST} catalogue associated with the cluster ($N_{\star}$). We employed \texttt{emcee} \citep{foreman-mackey+2013} to sample the posterior distribution of the model parameters. Further details on the likelihood function and the sampling procedure are provided in Appendix~\ref{app:struct}, together with the corner plot illustrating the marginalised one- and two-dimensional posterior distributions.
The lower panel of Figure~\ref{fig: density_map} illustrates the best-fit model, while Table~\ref{tab: total_tab} lists the 50th percentiles of the marginalised probability distribution functions of the model parameters, together with the 16th and 84th percentiles.

The structural properties of Gep~I obtained from our resolved star analysis are in good agreement with those reported in \cite{cusano+2016}, although these authors did not provide uncertainties and characterised the cluster using only integrated-light photometry. Our $\Rh$ estimate for the cluster is $0.7\arcsec$ larger, while we find an offset with respect to their centroid of the order of $1.7\arcsec$ and  $0.6\arcsec$ in right ascension and declination, respectively. In addition, from our fit, we recovered only an upper limit on the ellipticity and large uncertainties in the position angle (see Table~\ref{tab: total_tab}), consistent with the largely circular appearance of the cluster core in Figure~\ref{fig: density_map}. While there are some low-level hints of irregularity in the outer regions of Gep~I, deeper photometric observations are required to quantify this. 

\begin{table}
    \centering
    \renewcommand{\arraystretch}{1.75}
    \begin{tabular}{lr}
        \hline
        \textbf{Parameter} & \textbf{Value}  \\
        \hline
        $\alpha$ (deg) &  $7.54361^{+0.00017}_{-0.00017}$ \\    
        $\delta$ (deg) &   $46.85139^{+0.00017}_{-0.00020}$ \\
        $\Rh$ (asec)   &  $6.7^{+1.3}_{-1.0}$ \\
        $\Rh$ (parsec) &   $24^{+5}_{-4}$ \\
        $e$            &  $0.17^{+0.14}_{-0.11}$ \\
        $\phi$ (deg)   &  $60^{+60}_{-30}$  \\
        $N_{\star}$    &  $102 \pm 16$ \\
        \bottomrule
\end{tabular}
\caption{Derived structural properties for Gep~I by means of resolved HST-ACS/WFC photometry. The linear size is computed by assuming the distance to And~XXV that is measured in \citet{savino+2022}.}
\label{tab: total_tab}
\end{table}

 \begin{figure*}
    \centering
    \includegraphics[width=0.8\textwidth]{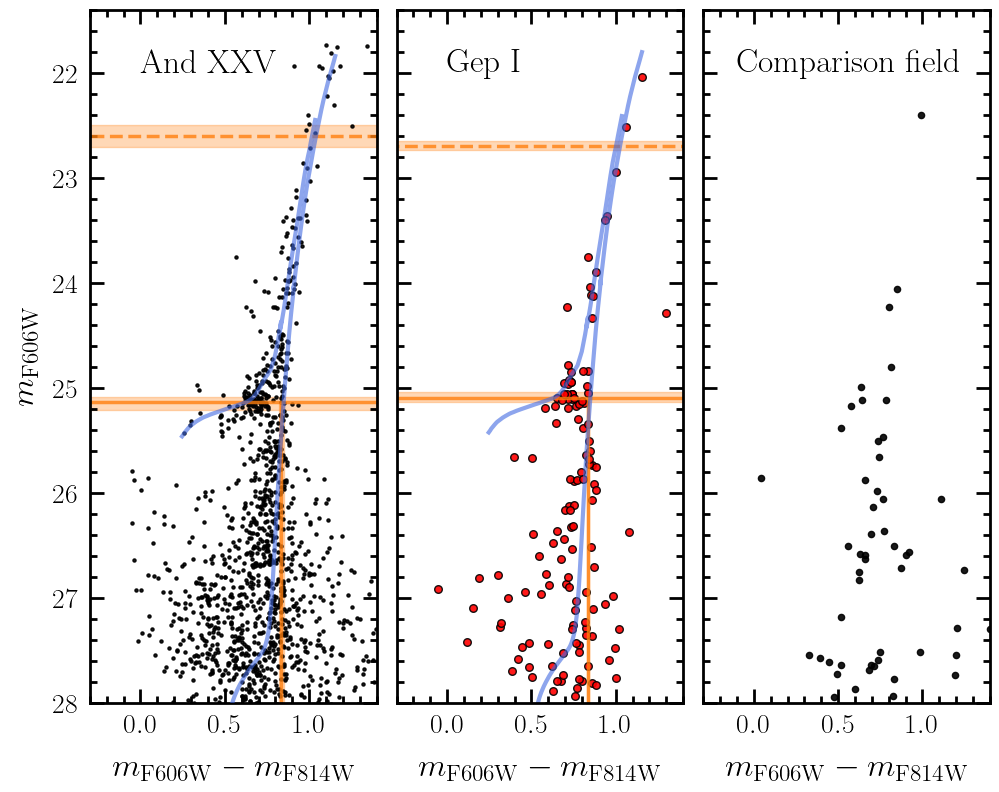}
    \caption{CMDs of stars passing the photometric quality cuts described in Sect.~\ref{sec: observation_reduction}. Left panel: CMD of stars located beyond $3\Rh$ from the cluster centre; we considered this sample as representative of the field populations of And~XXV. Central panel: CMD of stars inside $2\Rh$ from the cluster centre; we flagged these stars as Gep~I members. Right panel: CMD of stars within the same area as the central panel but centred on coordinates just outside $3\Rh$, serving as a local control field to evaluate background contamination in the cluster's CMD. In the left and central panels, the blue lines correspond to a 10 Gyr-old, $\alpha$-enhanced ($\rm [\alpha/Fe] = +0.4$) BaSTI isochrone. Adopted values of $\FeH$, $E(B-V)$ and the distance to And~XXV and Gep~I are the best-fit results from the CMD indices, as described in Sec.~\ref {sec: CMD_analysis} (see also Fig.~\ref{fig: fit_RGB_results}). These values are listed in Table~\ref{tab: cmd_fit}. The horizontal and vertical solid orange lines mark $V_{\rm HB}$ and $C_{\rm RGB}$, while the horizontal dashed orange line marks the derived value of $\Delta V_{0.2}$. Orange bands indicate the uncertainties in the respective measurements.}
\label{fig: CMD_LF}
\end{figure*}

\subsection{Colour-magnitude diagram analysis}
\label{sec: CMD_analysis}

Figure~\ref{fig: CMD_LF} presents the CMDs of stars from the culled photometric catalogue. The left panel displays stars located beyond $3\Rh$ from the cluster centroid, as determined in Sec.~\ref{sec: structure_fit}, while the central panel shows stars within $2\Rh$. This radius was chosen to maximise the number of stars likely associated with the cluster while maintaining a high probability of membership to Gep~I ($p > 0.75$), as determined by the fitting procedure described above. We consider these stars to be representative of the stellar cluster. As shown in Fig.~\ref{fig: CMD_LF}, stars within $2\Rh$ (red points) trace a tight, well-defined RGB and a populous red HB. We provide the comparison between the CMD of stars within only $1\Rh$ and those outside $3\Rh$ in Appendix~\ref{app:CMD_1Rh}.
The right panel of Figure~\ref{fig: CMD_LF} illustrates the CMD of a control field just outside $3\Rh$. The small number of field stars with magnitudes equal to or brighter than the HB ensures that background contamination is unlikely to affect the following analysis.

These CMDs represent a considerable improvement over the photometry presented in \cite{cusano+2016}, enabling, for the first time, a detailed comparison between the stellar population of the candidate cluster and that of the surrounding dwarf galaxy. 
To this aim, we followed \citet{mcgill+2025} by measuring a series of CMD indices -- namely, the magnitude of the HB ($V_{\rm HB}$), the corresponding RGB colour at the level of the HB ($C_{\rm RGB}$), and the slope of the upper RGB  -- and used the empirical relationships derived by Mackey et al. (2026, subm.) to translate these quantities into $\FeH$, reddening, and distance. The upper RGB slope is quantified by measuring the difference between $V_{\rm HB}$ and the magnitude of the RGB redward of $C_{RGB}$ by $\delta C  = 0.2$. %To quantify the upper RGB slope, a polynomial is fit anchored at ($C_{RGB}$, $V_{HB}$) and by the magnitude of the RGB redward of $C_{RGB}$ by $\delta C  = 0.2$ at the other.  
The upper RGB slope is used to infer the metallicity; from this, the intrinsic RGB colour could be determined, allowing the line-of-sight reddening, $E(B-V)$, to be measured via comparison to $C_{RGB}$. Finally, once the line-of-sight extinction is established, the distance is derived by comparing the observed HB magnitude, $V_{HB}$, with the intrinsic value expected for its metallicity.  

The empirical relationships between CMD indices and physical quantities are calibrated using 47 MW GCs from the ACS Globular Cluster Treasury survey \citep[GO-10775,][]{sarajedini2007} and a smaller extension program GO-11586, \citep[GO-11586,][]{Dotter2011}, and resultant metallicities are on the scale of \citet{carretta2009intrinsic-f89}. While the procedure is calibrated using old GCs and therefore specifically tailored for such systems, it has also proven successful when applied to sparse GCs \citep{mcgill+2025} and low-luminosity dSphs (Mackey et al.~2026, subm.). We apply this methodology to characterise the CMD of the candidate cluster and the surrounding field independently.

We first determined $V_{\rm HB}$ by identifying the peak in the luminosity function (LF) of stars within a colour range chosen to capture the HB. The LF was convolved with the Epanechnikov kernel from \texttt{scikit-learn} \citep{pedregosa+2011}, using a bandwidth of 0.055 mag, and the error on $V_{\rm HB}$ was defined as the interval over which the density falls to 0.7 of its peak value. To measure $C_{\rm RGB}$, we then selected RGB stars within $\pm0.3\,$mag about $V_{\rm HB}$. To detect the peak of this distribution, we applied the same kernel density estimate, this time with a bandwidth of 0.007 mag. Here, the error was set to the standard deviation of the distribution obtained from $\sim$1000 of Monte Carlo realisations, in which random Gaussian deviates about the HB magnitude were generated and a new $C_{\rm RGB}$ value was calculated each time. The measured values of $V_{\rm HB}$ and $C_{\rm RGB}$ are shown in Figure~\ref{fig: CMD_LF} by the horizontal and vertical orange solid lines, while the corresponding uncertainties are indicated by the shaded orange areas.
Next, we fitted a second-order polynomial to the bright RGB, anchoring it at ($C_{\rm RGB}$, $V_{\rm HB}$). We quantified the RGB curvature from the magnitude difference between $V_{\rm HB}$ and at a colour 0.2 mag redder than $C_{\rm RGB}$ ($\Delta V_{0.2}$; see dashed horizontal orange lines in Figure~\ref{fig: CMD_LF}). The associated error was estimated with the same Monte Carlo procedure described above.

\begin{figure}
    \centering
    \includegraphics[width=0.48\textwidth]{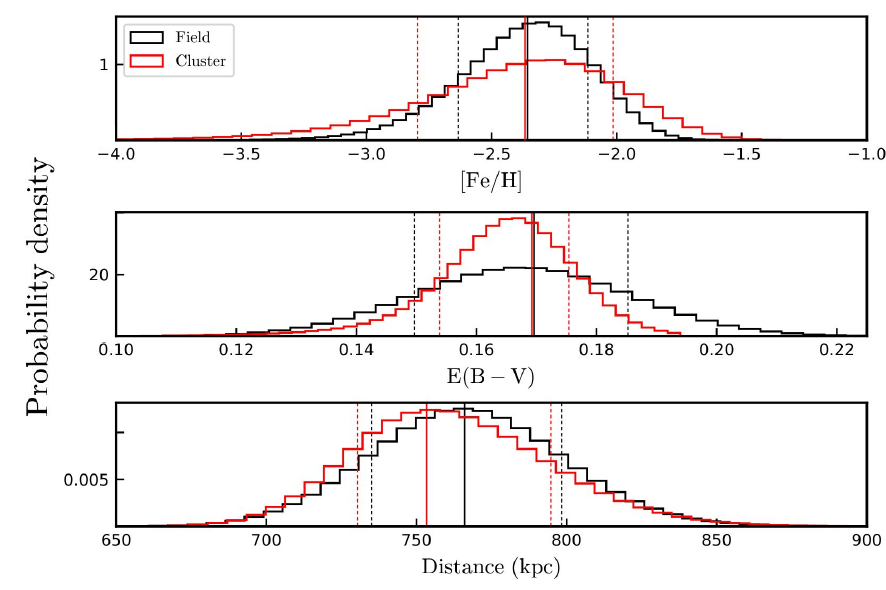}
    \caption{Probability density distributions resulting from the Monte Carlo method described in the text (see Sect.~\ref{sec: CMD_analysis}). Black and red histograms represent the results for And~XXV field stars and Gep~I stars, respectively. The derived values of $\FeH$, $E(B-V)$ and distance -- determined using the measured values of $V_{\rm HB}$, $C_{\rm RGB}$ and $\Delta V_{0.2}$ -- are indicated by the solid lines. The 16th and 84th percentiles of these distributions (dashed vertical lines) were used to determine the uncertainties on the measured quantities.}
\label{fig: fit_RGB_results}
\end{figure}

\begin{table*}
    \centering
    \renewcommand{\arraystretch}{1.75}
    \begin{tabular}{lccccccc}
        \hline
        \textbf{Target} & $V_{\rm HB}$  & $C_{\rm RGB}$  & $\Delta V_{0.2}$ & $\rm [Fe/H]$  &  $E(B-V)$  &  $D\,$(kpc)  \\
        \hline
        Gep~I  &  $25.10^{+0.03}_{-0.05}$  & $0.834^{+0.007}_{-0.007}$   &  $2.41^{+0.04}_{-0.04}$  &   $-2.4^{+0.3}_{-0.4}$   &  $0.169^{+0.006}_{-0.015}$    &  $753^{+41}_{-23}$   \\
        And~XXV  &  $25.14^{+0.07}_{-0.05}$  & $0.835^{+0.016}_{-0.029}$ &  $2.54^{+0.11}_{-0.11}$ &   $-2.4^{+0.2}_{-0.3}$   &  $0.170^{+0.016}_{-0.020}$      &  $766^{+32}_{-31}$    \\
        \bottomrule
\end{tabular}
\caption{Inferred CMD indices (HB magnitudes, corresponding RGB colour, magnitude difference between $V_{\rm HB}$ and the intersect with the best-fit polynomial), metallicity, reddening, and distance for Gep~I and And~XXV following the method described in Section~\ref{sec: CMD_analysis}. Stars within $2\Rh$ and outside $3\Rh$ from the cluster's centre were considered as representative of Gep~I and And~XXV, respectively.}\label{tab: cmd_fit}
\end{table*}

The derived index values are then fed into the empirical relationships provided by Mackey et al. (2026, subm.), which relate these fiducial measurements to the physical quantities of interest. We list these equations in Appendix~\ref{app:cal_rel}.
Figure~\ref{fig: fit_RGB_results} shows solid black and red lines indicating the inferred metallicity, reddening, and distance values of And~XXV field population and Gep~I, respectively. To estimate the uncertainty on these measurements, we performed 10,000 Monte Carlo extractions, generating Gaussian deviates around each index value with variances set by their derived uncertainties. For each extraction, we applied a randomly-drawn calibration equation using the relevant covariance matrix (see Mackey et al.~2026, subm.). The dashed lines in Figure~\ref{fig: fit_RGB_results} mark the 16th and 84th percentiles of these distributions, defining the measurement uncertainty. The results are also summarised in Table~\ref{tab: cmd_fit}. 

The best-fit values of $\rm [Fe/H]$, $E(B-V)$, and distance were used to compute the blue isochrones shown in Figure~\ref{fig: CMD_LF}. These isochrones correspond to a $10\,$Gyr old\footnote{The age has been chosen based on the star formation history of And~XXV derived by \cite{savino+2025}. Specifically, this is the time at which 90\% of the total star formation occurred.} and $\alpha$-enhanced ($\rm [\alpha/Fe] = +0.4$) BaSTI model \citep{hidalgo+2018,pietrinferni+2021}. As can be seen, these isochrones match the observed CMDs very well, validating the results of our fitting method (which is purely empirical). We therefore proceed to use the isochrone matching Gep~I's stellar population to estimate its luminosity and stellar mass. To do this, we generated a $10^6\,M_{\sun}$ synthetic stellar cluster population by sampling a Kroupa initial mass function with the python package \texttt{imf}\footnote{https://github.com/keflavich/imf}. We then rescaled it to match the number of stars observed above the detection limit in the $F606W$ band ($m_{\rm F606W} = 28$, see Fig.~\ref{fig: CMD_LF}). Uncertainties were propagated via Monte Carlo sampling of the posterior distributions of the fitted number of member stars, distance, and reddening.  The resulting absolute magnitude is $M_V=-4.5 \pm0.2\,$mag, while the corresponding stellar mass is $M_{\star}= 6.5\pm 1.3 \times10^3\,M_{\sun}$. We note that these estimates should be interpreted as lower limits, as we assume 100\% completeness for stars brighter than our detection threshold. Nonetheless, our estimated total magnitude is in reasonable agreement with the value previously reported in the literature from an integrated-light analysis ($M_V\sim -4.9\,$mag, \citealt{cusano+2016}).

Given their common line of sight, very similar reddening values are expected for And~XXV and Gep~I, and this is consistent with our measurements of 0.169 and 0.170 for And~XXV and Gep~I, respectively (see Table\,\ref{tab: cmd_fit}). The reddening value predicted by \citet{Schlegel1998}, as recalibrated by \citet{Schlafly2011}, for the And~XXV sightline is $E(B-V)=0.1$, which is slightly lower than our measurements. However, such a difference is not unexpected, given the relatively coarse spatial resolution ($\sim$6 arcmin) of the \citet{Schlegel1998} dust maps compared to the pencil-beam measurements made here, which are more affected by small-scale dust structure along the sight-line (see also McGill et al. 2026, subm.)

On the other hand, the derivation of very similar metallicities ($\FeH_{\rm And\,XXV} = -2.4^{+0.2}_{-0.3}\,$dex and $\FeH_{\rm Gep\,I} = -2.4^{+0.3}_{-0.4}\,$dex) and distances ($D_{\rm And\,XXV} = 766^{+32}_{-31}\,$kpc and $D_{\rm And\,XXV} = 753^{+41}_{-23}\,$kpc) is interesting, strongly supporting the physical association of Gep~I and the dwarf galaxy.  The $\FeH$ values derived from our photometry are very metal-poor, yet consistent with the only spectroscopically-determined value available for this galaxy: $\FeH = -1.9 \pm 0.1\,$dex (\citealt{collins+2013, charles+2023}). It is important to note, however, that direct comparison between these two measurements is difficult because they rely on different metallicity scales. In addition, our calibration relation is based on $\alpha$-enhanced Milky Way GCs, which may introduce metallicity offsets of about $0.3\,$dex when analysing solar-scaled systems (see Mackey et al. 2026, subm.). Regardless of the absolute metallicity, we can assert that both And~XXV and Gep~I are comparably rather metal-poor.  Furthermore, the line-of-sight distances measured for And~XXV and Gep~I are highly consistent,  and are in excellent agreement with the distance of And~XXV derived from RR-Lyrae stars \citep[$D = 751.6^{+25}_{-21}\,$kpc;][]{savino+2022}, as well as previous literature values obtained through alternative methods (see \citealt{richardson+2011,conn+2012,weisz+2019}). 

\section{Discussion }
\label{sec: discussion}

%On the basis of the results presented in Section~\ref{sec: CMD_analysis}, our analysis strongly encourages the association of Gep~I with the And~XXV dSph galaxy, although a sound confirmation of the membership will be possible only via kinematic data. We stress this result, since Gep~I has been included in several studies that empirically model the expected number and mass of GCs as a function of the host galaxy properties in the low-mass regime, where no consensus -- both observational and theoretical -- has been reached yet. On the one hand, such studies attempt to verify whether linear scaling relations, robustly observed for massive systems, can be extrapolated towards the dwarf-mass regime \citep{forbes+2018}; on the other hand, they test whether the observed rarity of GCs in dwarfs demands specific statistical count models \citep{eadie+2022, berek+2023, berek+2024}. Determining these empirical relations has a deep impact on the physical processes at the base of the formation and evolution of GCs implemented in cosmological hydrodynamical simulations (e.g., \citealt{bastian+2020, chen_gnedin2023}). %Ultimately, this reflects on our understanding of hierarchical assembly of galaxies, being dwarf systems the building blocks of such process.

\subsection{Gep~I as the solitary star cluster of And~XXV}

On the basis of the results presented in Section~\ref{sec: CMD_analysis}, our analysis strongly supports the physical association of Gep~I with the And~XXV dSph galaxy, although a definitive confirmation of membership will be possible only by measuring the radial velocity of the cluster.
The CMD of Gep~I closely resembles that of the main body of And~XXV, and we have shown that they share a common metallicity.
On the other hand, the current dataset does not allow us to reach a conclusion regarding the age similarity between the cluster and the dSph, as our photometry does not reach the main-sequence turn-off (MSTO). This limitation arises because only a fraction of the exposures obtained of And~XXV {\it HST} Treasury program captured Gep~I (see Sec.~\ref{sec: observation_reduction}). 

Curiously, the cluster exhibits a very red horizontal branch in spite of its low metallicity. While the HB morphology is known to be influenced by various parameters, such as mass-loss and He abundance, a common way to produce such a red HB at low metallicity is to assume a relatively young age \citep{hidalgo+2018, pietrinferni+2021}. Yet, the SFH of the And~XXV presented in \cite{savino+2025}, which is based on photometry reaching its old MSTO, indicates that the system is old and that it quenched $\sim$10 Gyr ago. Therefore, an age younger than this for Gep~I would be puzzling, suggesting that it formed later than the bulk of the stellar mass in And~XXV, while sharing the same metallicity.
Recently, using the same methodology applied in this study, \cite{mcgill+2025} and McGill et al. (2026, subm.) determined HB morphologies and $\FeH$ values for 48 GCs in the halo of M31, and 26 GCs in three Local Group dwarf galaxies, respectively. This work identified a small number of GCs in the halos of some of these systems which were characterised by low metallicities ($\FeH \leq -1.8$ dex) and red HBs, consistent with that seen in Gep~I. The authors argue that these peculiar GCs were likely recently accreted along with their now-destroyed low-mass companions.  In the case of M31, the evidence for this is particularly compelling as the clusters in question are associated with tidal debris streams in the halo \citep{mcgill+2025,mackey+2019}. 

Within the Milky Way, the GCs most similar to Gep~I are Crater \citep{weisz+2016} and AM-1 \citet{2011ApJ73874D}.  These are both characterised by very red HBs, low metallicities (Crater: \citep[$\lbrack \text{Fe/H} \rbrack = -1.68$,][]{kirby+2015, bonifacio+2015}, AM-1: \citep[$\lbrack \text{Fe/H} \rbrack = -1.84$,][]{carretta2009intrinsic-f89} and large sizes \citep[$\sim 15-20$~pc,][]{pace2025}. MSTO age determinations suggest that AM-1 is slightly younger \citep[][$\sim11$ Gyr]{dotter+2008} than typical outer-halo GCs in the Milky Way, while Crater is substantially younger \citep[][$\sim7.5$ Gyr]{weisz+2016}. Both these GCs lie in the far outer halo ($\gtrsim120$~kpc) of the Miky Way, and both are suspected to be accreted, with AM-1 specifically tagged, along with one or two other clusters on similarly high-energy retrograde orbits, to the Elqui stream \citep{massari2025origin-4da,deleo+2026}. Crater is not associated with any of the major accretion events that contributed to the Galactic stellar halo \citep{callingham+2022,massari2025origin-4da} but it has been proposed to be dynamically linked to a group of four Milky Way satellites, known as the Crater-Leo group, which includes Leo~II, Leo~IV, and Leo~V. The infall of this group likely occurred within the past 6 Gyr \citep{julio+2024}. Within these satellites, the dwarf spheroidal galaxy Leo~II, whose luminosity ($M_V = -9.8$; \citealt{irwin+1995}) is comparable to that of And~XXV, is considered the probable progenitor of Crater due to its similar metallicity.  Gep~I would appear to be an analogue of these outer-halo GCs, but one which is still residing within its host dwarf galaxy.

On the other hand, the striking similarity between the CMD of Gep~I and that of And~XXV field stars makes this stellar overdensity a compelling candidate for investigating the presence of a dark matter subhalo. Small-scale dark substructures are predicted by the $\Lambda$CDM cosmology, and \cite{penarrubia+2024} argued that these subhalos can temporarily capture field stars as they orbit within the dark matter-dominated potential wells of dSph galaxies. Stellar systems formed through such interactions are expected to closely resemble the surrounding field dSph population, have extended sizes for their luminosities, and possess high mass-to-light ratios. While the photometric and structural properties of Gep~I are consistent with the simulations of \cite{penarrubia+2024}, the absence of kinematical data precludes assessment of whether the cluster exhibits an anomalously high velocity dispersion.

\subsection{Comparing the properties of the star clusters in And~XXV and Eri~II}

In this section, we compare the observed properties of Gep~I with those of the stellar cluster in Eridanus~II (Eri~II), which represents the only other well-studied example of a very low-mass dwarf hosting a solitary GC (\citealt{crojevic+2016,zoutendijk+2020, simon+2021, weisz+2023}).
Relative to the majority of GCs orbiting in the Milky Way (see \citealt{baumgardt_hilker2018}), these dwarf-hosted clusters are notably extended and faint. Gep~I appears larger ($\Rh= 24\,$pc) and brighter ($M_V=-4.5$) than Eri~II's cluster, which is measured to have $\Rh= 15\,$pc and $M_V = -3.4$ (\citealt{simon+2021, weisz+2023}). This may reflect different birth conditions within their hosts, as And~XXV is more luminous ($M_V = - 9.1$; \citealt{savino+2025}) than Eri~II ($M_V = - 7.1$; \citealt{crojevic+2016}), or it may result from different tidal fields and mass-loss rates. As noted in \cite{simon+2021}, the large ellipticity of Eri~II's cluster ($e=0.31^{+0.05}_{-0.06}$), and its perfect alignment with the orientation of the parent galaxy, hint at an advanced stage of tidal-stretching. 
This would justify the lower stellar mass for Eri~II's cluster; yet, no signs of tidal stripping have been found around this system. In contrast, such a large elongation and well-defined position angle are not observed in Gep~I, although deeper photometric observations are required to make this conclusion more robust. 

Notably, the two clusters are characterised by a distinct offset from their host galaxy centres. While Eri~II's cluster is observed at a very small projected distance from the galaxy centre ($23\pm3\,$pc, \citealt{simon+2021}), we find that Gep~I is offset from And~XXV's centre by $138\pm55\,$pc. This offset was determined by measuring the angular distance between the cluster centre, as determined in Sec.~\ref {sec: structure_fit}, and the galaxy centre, as reported by \cite{martin+2016}. For consistency, the coordinates from \cite{martin+2016} were transformed to our astrometric frame using stars common to both catalogues. The resulting value is $\theta = 38\arcsec\pm15\arcsec$. This measurement and its associated uncertainty were estimated by sampling Gaussian distributions with variances set by the measured errors in the central coordinates, and by considering the 50th, 16th, and 84th percentiles of the resulting posterior distribution. The angular offset was then converted to the linear projected distance using the And~XXV distance reported in \cite{savino+2022}. 

The observed properties of the star clusters in Eri~II and Gep~I provide valuable constraints for N-body simulations that aim to use their longevity to make inferences about the DM distribution in their host galaxies. 
Several simulations have been carried out focusing on Eri~II's cluster \citep{amorisco2017,contenta+2018,webb&vesperini2018,orkney+2022}. All of these studies struggle to reconcile the present-day properties of the cluster with a cuspy DM profile, the halo configuration predicted by structure formation simulations in a $\Lambda$CDM cosmology \citep{navarro+1996}. Within such a halo, only highly specific conditions would allow the cluster to survive for several Gyrs. In contrast, a scenario involving a large central dark matter core would naturally explain the cluster's survival, size, ellipticity, and projected position.  In light of our improved characterisation of Gep~I, similar studies applied to And~XXV would be extremely informative. Indeed, kinematic analyses of And~XXV's field stars have suggested an unusually low central dark matter density in this galaxy, which could possibly be explained by past tidal interactions with M31 (see \citealt{collins+2013, charles+2023}). While substantial uncertainties in the inferred dark matter profile preclude definitive conclusions, the survival of a low-mass, extended star cluster deep in its potential well may bring further insight.

%tune assembly histories and sub-grid baryonic physics implemented in cosmological simulations. 
%Nevertheless, it is tantalising to note that, although characterised by low metallicity, the cluster exhibits a very red HB, in contrast to the blue HB observed in Eri~II

\section{Summary and conclusions}
\label{sec: summary_conclusions}

In this paper, we have characterised the candidate star cluster of And~XXV, called Gep~I, by means of deep archival {\it HST}-ACS/WFC observations. This is the first detailed resolved star analysis of this object, and our resulting CMD reaches to $\gtrsim2$ magnitudes below the HB. We demonstrate the existence of a resolved star overdensity at the position of Gep~I, and we employ new empirical relations to determine $\FeH$, line-of-sight reddening and distance to both the cluster and the surrounding And~XXV field. We find excellent agreement in all properties, strongly supporting the association of the cluster with the dSph, as first proposed by \cite{cusano+2016}, and alleviating concerns about the chance projection of an M31 halo GC. And~XXV is therefore one of the few known very low-mass galaxies ($M_{\star} < 10^6\,M_{\sun}$) hosting a GC, along with the ultra-faint dwarf galaxy Eri~II. 

We find Gep~I to be very extended ($\Rh=24^{+5}_{-4}$ pc), faint ($M_V = -4.5 \pm 0.2$), and located at a projected distance of only $138\pm55\,$pc from the galaxy centre. From the CMD, we measure a photometric metallicity of $\FeH = -2.4^{+0.3}_{-0.4}$ dex, a line-of-sight reddening of $E(B-V) = 0.17$, and distance $D = 753^{+41}_{-23}\,$kpc.  Although it is metal-poor, the cluster exhibits a very red HB morphology.  This combination of parameters has recently been found in a small number of GCs throughout the Local Group, including in the halo of M31 and the outskirts of the dIrr galaxy NGC\,6822.  In all of these cases, the clusters are hypothesised to be comparatively young and recently accreted from low-mass galaxies \citep{mcgill+2025}. Although slightly more metal-poor, Gep~I also shares similarities with the Milky Way outer-halo GCs Crater and AM-1, both of which also have a strong suggested accretion origin and younger ages than the dominant old halo population. Gep~I is a likely analogue of these outer-halo GCs, but one which is still residing within its host dwarf galaxy halo.  An age estimate for Gep~I would establish whether it formed along with, or later than, the bulk of And~XXV's stellar mass, which was 90 per cent in place by 10 Gyr (see \citealt{savino+2025}). 

With the improved characterisation of Gep~I, it will be possible to construct new models to explore whether such a cluster could survive in the inferred potential well of And~XXV, shedding light on the nature of dark matter, the properties of low-mass haloes, and how they are affected by galaxy formation. 
Furthermore, a velocity dispersion measurement will clarify if Gep~I represents a genuine long-lived star cluster, or is a transient stellar overdensity resulting from stars captured by a dark matter subhalo orbiting within the And~XXV's potential well.

\section*{Acknowledgements}
We thank the referee, Michelle Collins, for the helpful comments that improved the paper.
We thank David Behrendt for his assistance during the early stages of this project, and Nicolas Martin for useful discussions. CC and AMNF are supported by the Science and Technology Facilities Council [grant number ST/Y001281/1]. AMNF also acknowledges support from UK Research and Innovation (UKRI) under the UK government’s Horizon Europe funding guarantee [grant number EP/Z534353/1]. GM acknowledges funding from the Bell Burnell Graduate Scholarship Fund [grant number BB0027]. Based on observations collected with the NASA/ESA HST under the program GO-15902, and at the Gemini South Observatory under the program GS-2007B-Q-23.
This research made use of the following python libraries: \texttt{Matplotlib} \citep{matplotlib}, \texttt{Numpy} \citep{numpy}, \texttt{Scipy} \citep{scipy}, \texttt{scikit-learn} \citep{pedregosa+2011}, \texttt{Astropy} \citep{astropy}.
%%%%%%%%%%%%%%%%%%%%%%%%%%%%%%%%%%%%%%%%%%%%%%%%%%
\section*{Data Availability}

The data used in this analysis is available from the Mikulski Archive for Space Telescopes (MAST) with the programme ID GO-15902.
The photometric catalogue used for this article will be shared upon a reasonable request to the corresponding author.

%%%%%%%%%%%%%%%%%%%% REFERENCES %%%%%%%%%%%%%%%%%%

% The best way to enter references is to use BibTeX:

\bibliographystyle{mnras}
\bibliography{mnras_biblio} % if your bibtex file is called example.bib

\begin{appendix}
\section{Structural fit}
\label{app:struct}

This appendix provides the details of the likelihood function and the Markov Chain Monte Carlo sampling adopted for the structural analysis. The likelihood is defined as

\begin{equation}\label{for:likl}
    \mathcal{L}(\bbtheta\,|\, X_j,Y_j)=
    \prod_{j=1}^{N_\star}
    \mathcal{P}(X_j,Y_j\,|\,\bbtheta),
\end{equation}
where $\mathcal{P}$ is the model probability density, $(X_j,Y_j)$ are the sky coordinates of the $j$-th star and $\bbtheta$ denotes the set of model parameters. The model probability density is composed of the star cluster and a foreground component,
\begin{equation}\begin{split}\label{for:prob}
    \mathcal{P}(X,Y \,|\, \bbtheta)= &
    w\,
    \frac{\Sigma_{\rm cl}(X,Y\,|\,\bbtheta)}
    {\int_{\Area}\Sigma_{\rm cl}(X,Y\,|\,\bbtheta)\dd X\dd Y} \\ & 
    +(1-w)\,
    \frac{\Sigma_{\rm bg}(X,Y\,|\,\bbtheta)}
    {\int_{\Area}\Sigma_{\rm bg}(X,Y\,|\,\bbtheta)\dd X\dd Y},
\end{split}\end{equation}
where $0<w<1$ is the fraction of stars associated with the cluster ($\Sigma_{\rm cl}$), $\Sigma_{\rm bg}$ is the background component, and the integrals in the denominators ensure that the two terms are normalised to unity over the observed sky region ($\Area$). The cluster component is described by the flattened \cite{plummer1911} profile
\begin{equation}
    \Sigma_{\rm cl}(m)=\frac{1}{(1-e)\pi\Rh^2}
    \left(1+\frac{m^2}{\Rh^2}\right)^{-2},
\end{equation}
where $\Rh$ is the projected half-light radius and the elliptical radius is defined as
\begin{equation}
\begin{split}
    m^2 = & \left[\frac{(X-X_0)\cos\phi-(Y-Y_0)\sin\phi} {1-e}\right]^2 \\ &+\left[(X-X_0)\sin\phi+(Y-Y_0)\cos\phi\right]^2.
\end{split}
\end{equation}
In the above equations, $\phi$ is the position angle, $e\equiv1-b/a$ is the ellipticity, with $b$ and $a$ the semi-minor and semi-major axes, respectively. The background component is defined as
\begin{equation}\begin{split}
    & \Sigma_{\rm bg}(X,Y)=1+\rho\left(X\cos\theta+Y\sin\theta\right), \\ & \Sigma_{\rm bg}(X,Y)>0, \quad \forall(X,Y)\in\Area.
\end{split}\end{equation}
In the above equation, $\rho$ is proportional to the gradient amplitude and $\theta$ is the direction of the gradient.

\begin{figure*}
    \centering
    \includegraphics[width=1\linewidth]{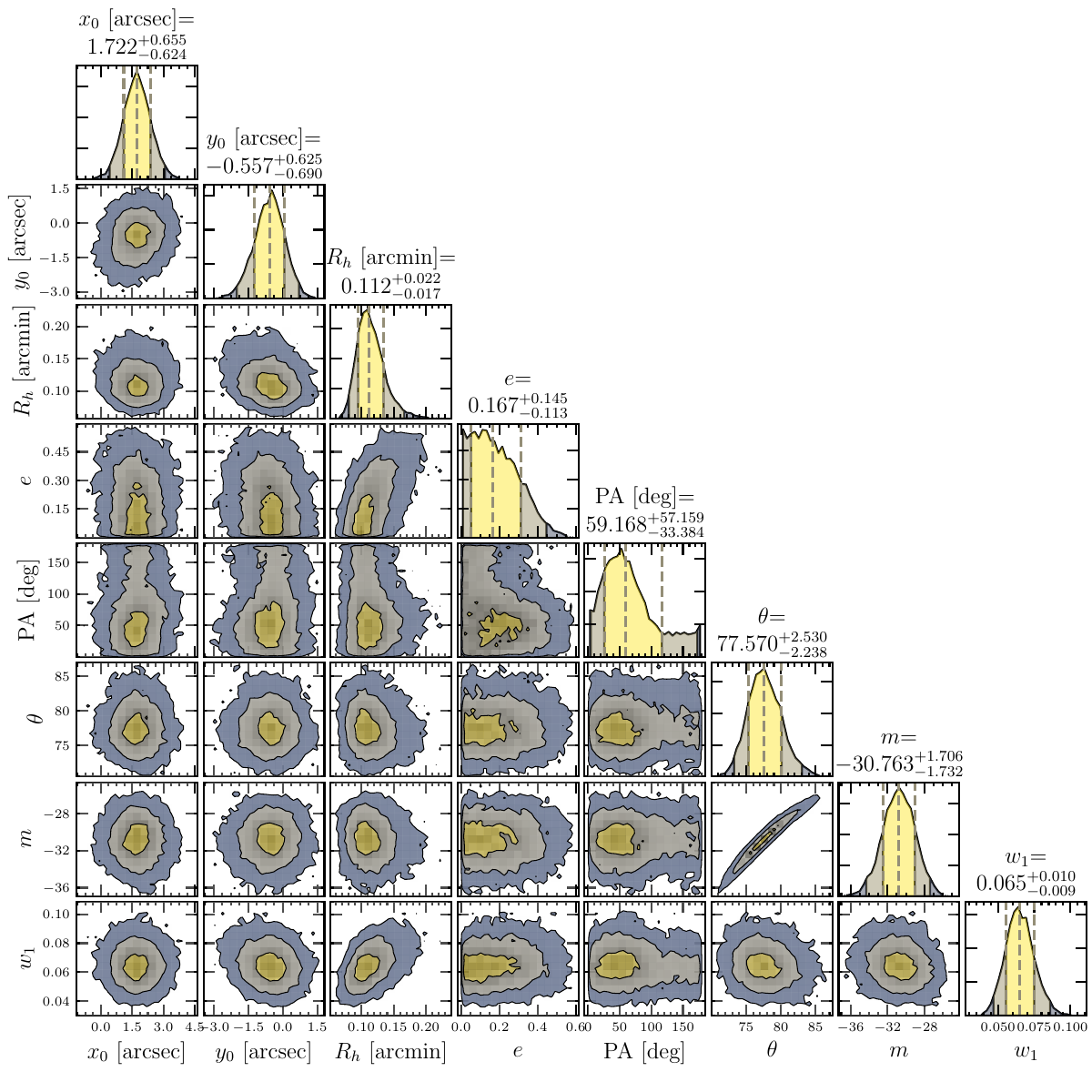}
    \caption{Marginalised one- and two-dimensional posterior distributions of the model's free parameters. The contours in the two-dimensional marginalised distributions enclose, respectively, 68\%, 95\%, and 99\% of the total probability. In the one-dimensional marginalised distributions, the grey lines indicate the 16th, 50th and 84th percentiles.}
    \label{fig:corner_plot}
\end{figure*}

Posterior distributions are sampled with the affine-invariant ensemble sampler implemented in \texttt{emcee}. We adopt 48 walkers evolved for 7000 steps and discard an initial burn-in phase of 2000 steps. Walkers that fail to converge, as diagnosed by the behaviour of the log-posterior, are removed. The adopted parameter values correspond to the medians of the posterior distributions, while uncertainties are computed from the 16th--84th (1$\sigma$). Figure~\ref{fig:corner_plot} shows the marginalised one- and two-dimensional posterior distributions on the model's free parameters.

\section{Additional colour-magnitude diagrams}
\label{app:CMD_1Rh}
\begin{figure}
    \centering
    \includegraphics[width=0.9\linewidth]{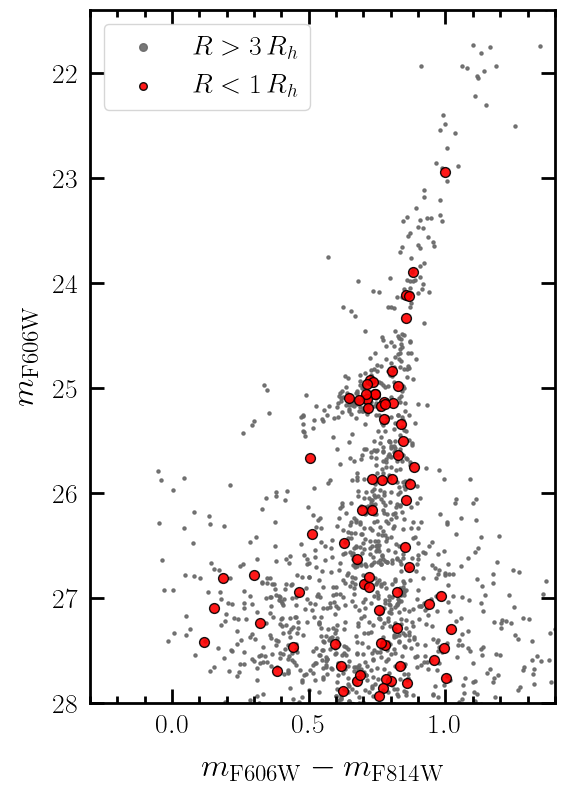}
    \caption{CMDs of stars passing the photometric quality cuts described in Sect.~\ref{sec: observation_reduction}. Grey points represent stars located outside $3\Rh$ from the centre of Gep~I, taken as representative of the And~XXV field population, while red markers represent stars within $1\Rh$.}
    \label{fig: CMD_1Rh}
\end{figure}

To underscore the similarity between the CMD of Gep~I and that of And~XXV, we show in Figure~\ref{fig: CMD_1Rh} the comparison between stars within $1\Rh$ (red markers), which have a probability of membership to Gep~I higher than 0.95, and outside $3\Rh$ (grey dots). While the red points are optimal for a qualitative comparison with And~XXV field stars, their low statistics hamper convergence of our adopted fitting methodology (see Section~\ref{sec: CMD_analysis}).

\section{empirical relationships}
\label{app:cal_rel}
Below, we list the relevant calibration relationships used in this work. These are based on a new empirical methodology described in Mackey et al. (2026, subm.), and employed in \cite{mcgill+2025} and McGill et al. (2026, subm.). These calibration relationships are tailored for ancient simple stellar populations observed with the ACS/WFC camera aboard {\it HST} using the filters F606W and F814W.

By exploiting the values of $V_{\rm HB}$, $C_{\rm RGB}$ and $\Delta V_{0.2}$ measured in Section~\ref{sec: CMD_analysis} (see Table~\ref{tab: cmd_fit}), we first derived metallicity values with the following expression:

\begin{equation}
\begin{split}
     {\rm [Fe/H]} & =  - 2.1442885(\Delta V_{0.2})^3 + 10.6949430(\Delta V_{0.2})^2  \\ &  - 18.5782241(\Delta V_{0.2}) + 10.3015266.
\end{split}
\end{equation}

Consequently, it is possible to compute the intrinsic colour of the RGB at the HB level via 

\begin{equation}
\begin{split}
     C_{0, \rm RGB} & =   0.0306324 \, \rm [Fe/H]^2 + 0.1482682 \, [Fe/H]  \\ & + 0.8539399,
\end{split}
\end{equation}

\noindent which allows the determination of $E(B-V)$ through the following equation:

\begin{equation}
    E(B-V) = (C_{\rm RGB} - C_{0,RGB}) / 0.945 .
\end{equation}

To derive the above equation, we considered the reddening coefficient from \cite{Schlafly2011}. Specifically:

\begin{equation}
\begin{split}
 &   A_{m_{\rm F606W}} = 2.471 \, E(B-V) \\ & A_{m_{\rm F814W}} = 1.526 \, E(B-V) \\ & E(m_{\rm F606W} - m_{\rm F814W}) = 0.945 \, E(B-V) .
\end{split}
\end{equation}

Finally, Mackey et al. (2026, subm.) presented a new relationship, based on GC distances reported in \cite{baumgardt_vasiliev2021}, relating metallicity and intrinsic luminosity of the HB:
\begin{equation}
    M_{\rm V, HB} = 0.1954416\, \rm [Fe/H] + 0.7572936 , 
\end{equation}

\noindent thanks to which it is possible to compute the distance in kpc through the following expression:
 
\begin{equation}
    {\rm log_{10}} D = 0.2\,(V_{\rm HB} - M_{\rm V, HB} - 2.471\,E(B-V) - 10) .
\end{equation}

\end{appendix}

% Alternatively you could enter them by hand, like this:
% This method is tedious and prone to error if you have lots of references
%\begin{thebibliography}{99}
%\bibitem[\protect\citeauthoryear{Author}{2012}]{Author2012}
%Author A.~N., 2013, Journal of Improbable Astronomy, 1, 1
%\bibitem[\protect\citeauthoryear{Others}{2013}]{Others2013}
%Others S., 2012, Journal of Interesting Stuff, 17, 198
%\end{thebibliography}

%%%%%%%%%%%%%%%%%%%%%%%%%%%%%%%%%%%%%%%%%%%%%%%%%%

%%%%%%%%%%%%%%%%% APPENDICES %%%%%%%%%%%%%%%%%%%%%

%\appendix

%\section{Some extra material}

%If you want to present additional material which would interrupt the flow of the main paper,
%it can be placed in an Appendix which appears after the list of references.

%%%%%%%%%%%%%%%%%%%%%%%%%%%%%%%%%%%%%%%%%%%%%%%%%%

% Don't change these lines
\bsp	% typesetting comment
\label{lastpage}
\end{document}